\documentclass[journal]{IEEEtran}
\usepackage{tabularx,booktabs}
\usepackage{ifpdf}
 \ifpdf
 \else
 \fi
\usepackage{multirow} 
\usepackage{booktabs}
\usepackage{threeparttable}
\usepackage{makecell}

\usepackage{cite} 
\usepackage{url} 
\usepackage{graphicx}
\usepackage{float}

\ifCLASSOPTIONcompsoc
 \usepackage[caption=false,font=normalsize,labelfont=sf,textfont=sf]{subfig}
\else
 \usepackage[caption=false,font=footnotesize]{subfig}
\fi
\usepackage{epstopdf}
\usepackage{array}
\usepackage{amssymb}
\usepackage{amsmath}
\usepackage{bm} 
\usepackage{comment}
\usepackage{color}

\makeatletter
\let\NAT@parse\undefined
\makeatother
\usepackage{hyperref}  

\begin{document}

\markboth{Journal of \LaTeX\ Class Files}%
{Shell \MakeLowercase{\textit{et al.}}: Bare Demo of IEEEtran.cls for IEEE Journals}

\title{Fast Transient Stability Prediction Using Grid-informed Temporal and Topological Embedding Deep Neural Network}
\author{Peiyuan~Sun, Long~Huo, Siyuan~Liang, and Xin~Chen
}

\date{2022.01.14}

\maketitle


\begin{abstract}
	Transient stability prediction is critically essential to the fast online assessment and maintaining the stable operation in power systems. The wide deployment of phasor measurement units (PMUs) promotes the development of data-driven approaches for transient stability assessment. This paper proposes the temporal and topological embedding deep neural network (TTEDNN) model to forecast transient stability with the early transient dynamics. The TTEDNN model can accurately and efficiently predict the transient stability by extracting the temporal and topological features from the time-series data of the early transient dynamics. The grid-informed adjacency matrix is used to incorporate the power grid structural and electrical parameter information. The transient dynamics simulation environments under the single-node and multiple-node perturbations are used to test the performance of the TTEDNN model for the IEEE 39-bus and IEEE 118-bus power systems. The results show that the TTEDNN model has the best and most robust prediction performance. Furthermore, the TTEDNN model also demonstrates the transfer capability to predict the transient stability in the more complicated transient dynamics simulation environments.
\end{abstract}


\IEEEpeerreviewmaketitle

\section{Introduction}

\IEEEPARstart{D}{igitalization} plays an essential role in the modernization of the power systems. The wide deployment of phasor
measurement units (PMUs) enable the data collection on wide-area power systems and help engineers in utilizing large volumes of data, analyzing dynamic events in the grid, and the data-driven real-time transient stability prediction \cite{en14113148, 8403460}. 
Furthermore, the uncertainties of decentralized renewable energies demand a fast online assessment of transient stability \cite{23333}.
Currently, the prediction methods of transient stability
can be classified into two categories, $i.e$., the model-driven
methods and the data-driven methods\cite{zhang2021critical}.

One of the commonly used model-driven methods is the
laborious Time-domain Simulation (TDS) based on high-
dimensional nonlinear differential-algebraic equations (DAEs)
that express the dynamics of power systems \cite{basin_stability_for_generator}. TDS is time-consuming 
since it demands the whole state trajectories to reveal the system stability. Although people have proposed different approaches to accelerate the TDS process, such as parallel computing \cite{basin_stability_for_generator}, advanced hardware \cite{zhang2019new}, $etc.$, huge computation
resources are still required to handle the increasing complexity
of power systems and diverse operational scenarios. The
Lyapunov functions family's model-driven method is used for an analytical approach for stability assessment in power
systems \cite{7106572, 7488976}. Unfortunately, finding a Lyapunov function to accurately evaluate the transient stability of power systems has been proved to be very difficult \cite{anghel2013algorithmic}.

Recently, the data-driven methods, especially the deep learning approaches, attracted a lot of research interests on predicting transient stability in power systems \cite{8403460, 8627948, en14113148, 7932885}. Compared with the model-driven methods, deep learning performance does not rely on the power systems' prior knowledge and model details. Furthermore, the strong generalization ability and the nature of offline training and online diagnosis pattern of deep learning provide great potentials to meet the high accuracy and fast online requirements in practical applications \cite{real_time}. 

Among the existing deep learning models, Convolution Neural network (CNN) has made significant achievements in many fields \cite{VGG, ResNet}, including transient stability prediction in power systems. For example, Hou et al. \cite{8403460} proposed a power system transient stability fast evaluation model based on CNN and the voltage phasor complex plane image. Rong Yan et al. \cite{8627948} designed cascaded CNNs to capture data from different TDS time intervals, extract features, predict stability probability, and determine TDS termination. Besides, many other deep learning models are also used for predicting transient stability, such as auto-encoder \cite{en14113148}, Long Short-Term Memory (LSTM) network \cite{7932885}, and Generative Adversarial Network (GAN) \cite{han2021imbalanced}. However, the architectures of the above deep learning models lack proper interpretability with the spatial correlations of power systems, given that, essentially, power systems are complex dynamical networks. Therefore, effectively using the important information of power network structures in deep learning remains a challenging problem. 

The recently developed Graph Neural Network (GNN) is a promising deep learning model to extract features of the spatial correlations of power systems since GNN can naturally map the power network structure into its neural network architecture. As one of the GNN family, Graph Convolution Network (GCN) \cite{kipf2017semisupervised} combines topological structure with convolution algorithm and has been proved to be extremely powerful for the complex dynamical network analysis such as traffic prediction \cite{2017arXiv170904875Y} and action recognition \cite{yan2018spatial}. GCN demonstrates good classification and prediction capability with the graph-structured data in power systems \cite{A_Comprehensive_Survey_on_Graph, A_Review_of_Graph}. For example, Yuxiao Liu et al. \cite{9369017} developed an interpretable GCN to guide cascading failure search efficiently. Nevertheless, the GCN is not adept at capturing the sequential characteristics, i.e., the temporal information of time series of power system dynamics. Additional techniques are needed to extract features from the time-domain of power system transient dynamics. For sequence modeling \cite{DL}, the convolutional technique has been developed extensively in recent works and outperformed the baseline of well-known recurrent network architectures for sequence modeling tasks \cite{oord2016wavenet}. As one of the convolutional technique-based recurrent architectures, Temporal Convolutional Network \cite{bai2018empirical}, also known as TCN, has been utilized for time-series predictions in power systems, demonstrating powerful memory ability \cite{tang2022short, li2022multi}. In this paper, the temporal and topological embedding deep neural network (TTEDNN) model is proposed combining GCN and TCN to capture the spatio-temporal features of transient dynamics in power systems.

Generally, the main contributions of this paper are as follows:

\begin{itemize}
	\item [1)] The TTEDNN model is proposed to predict the transient
	stability by the temporal and spatial features extracted from the time-series data of the early transient dynamics. The grid-informed adjacency matrix is used to incorporate the power grid structural and electrical parameter information;
	\item [2)] The performance of the TTEDNN model to predict the transient
	stability is highly efficient and accurate compared to existing deep learning
	methods and traditional TDS method;
	\item [3)] The predictive capability of the TTEDNN model is robust. The
	TTEDN model trained with the second-order swing equations demonstrates good performance to predict the transient stability in more complicated transient dynamics simulation environments, $i.e.$, the higher-order (11$th$ order) power system model.
\end{itemize}

The rest of this paper is organized as follows.
Section~\ref{section:dynamic}
introduces simulation environment of transient dynamics.
Section~\ref{section:GCN} proposes the architecture of the TTEDNN model.
Case studies are given in Section~\ref{section:case}.
The transfer predictive capability is investigated in Section~\ref{section:transfer}.
The conclusion remarks are drawn in Section~\ref{section:conclude}.

\section{Simulation Environment of Transient Dynamics} 
\label{section:dynamic}
In this section, the power system models and perturbations of the transient dynamics simulation environment are described as follows.

\subsection{Models} 
The simulation environment of transient dynamics is set up for the data
generation to train and test the TTEDNN model.
The topology of a connected power system can be modelled as a weighted
undirected graph $G=(\mathcal{V},\mathcal{E},\mathbf{Y})$,
where $\mathcal{V}$, $\mathcal{E}$ and $\mathbf{Y} \in\mathbb{C}^{N \times N}$
is the node set, edge set and the symmetric nodal admittance matrix,
respectively.
In graph $G$, nodes represent the generators or loads behind the buses
and the edges represent the transmission lines.
The notations of bus/node and transmission line/edge are used interchangeably.
The number of nodes and edges are indicated by $N$ and $E$, respectively. The swing equations, \textit{a.k.a.}, the second-order Kuramoto oscillators in
physical community~\cite{rodrigues2016kuramoto}, are used in the simulation environment, which have been proved to be suitable to
model the transient dynamics of power systems \cite{Analysis_of_a_power_grid_using, Synchronization_assessment_in}:

\begin{equation}
	I_i \omega_{syn} \dot{\omega}_{i} + D_i \omega_{i} = P_{m,i}-\sum_{i \neq j} P^{MAX}_{ij} \sin (\delta_{i} - \delta_{j})
	\label{eq:swing_equation_1}
\end{equation}
where $\delta_i \in \mathbb{S}^1$ and $\omega_i \in \mathbb{R}^1$ are phase angle and angular frequency of node $i \in \mathcal{V}$ with respect to
power system rated frequency $\omega_{syn}$,
$P_{m,i}$ is the mechanical power injection, $P_{m,i} \geqslant 0$ for
generator node and $P_{m,i} \leqslant  0$ for load node, $I_i > 0$ is the
cumulative moment of inertia and $D_i > 0$ is the damping coefficient.
For the edge $\{i,j\} \in \mathcal{E}$ between node $i$ and node $j$,
$P^{MAX}_{ij}=|V_i|\cdot |V_j| \cdot \Im (Y_{ij})$ is the maximum real
power transfer capacity, where $Y_{ij}$ denotes the element of
$\mathbf{Y}$, $|V_i|$ and $|V_j|$ are the voltage amplitude of node
$i$ and node $j$.
Suppose the power system is lossless, only the imaginary part of admittance
$\Im (Y_{ij})$ is considered.

Dividing Eq.~\ref{eq:swing_equation_1} by $I_i \omega_{syn}$ and define
$\alpha_i=D_i/I_i\omega_{syn}$, $P_i=P_{m,i}/I_i\omega_{syn}$ and
$K_{ij}=P^{MAX}_{ij}/I_i\omega_{syn}$, the second-order power system model can be described as~\cite{dorfler2013synchronization}

\begin{align}
	\left\{
	\begin{array}{rl}
		\dot{\delta}_{i} = & \omega_{i}                                                                       \\
		\dot{\omega}_{i} = & -\alpha_i \omega_{i}+P_{i}+\sum_{i \neq j} K_{ij} \sin (\delta_{j} - \delta_{i}).
	\end{array}
	\right.
	\label{eq:swing_equation}
\end{align}

A higher order ($11th$ order) power system model derived from the PST toolbox\cite{PST} is also considered in this paper, see reference book \cite{PSTbook} for modelling details.

In this paper, we determine the transient stability in terms of synchronization in power systems \cite{dorfler2012synchronization}. For $\gamma \in [0, \pi/2]$, let
$\overline{\Delta}_G(\gamma) \subset \mathbb{T}^n$ be closed set of angle
arrays $(\delta_1,...,\delta_n)$ with the property
$|\delta_i-\delta_j|\leq\gamma$ for $\{i,j\}\in\mathcal{E}$.
A solution $(\delta, \omega):\mathbb{R}_{\geq 0} \rightarrow (\mathbb{T}^n,\mathbb{R}^n)$
to the power system model of a given initial state
in Eq.~\ref{eq:swing_equation} is said to be synchronized if
$\delta(0) \in \overline{\Delta}_G (\gamma)$, $\delta(t)=\delta(0)$
and $\omega(t)=0_{n}$ for all $t \geq 0$.
In other words, here, synchronized trajectories have the properties of
\textit{frequency synchronization}, that is, all oscillators rotate with
the same synchronization frequency $\omega_{syn}$ and all their phases
belong to the set $\overline{\Delta}_{G}(\gamma)$~\cite{Synchronization_assessment_in}.

\subsection{Perturbations}

The perturbations in power systems have varieties of sources including
load variations, market trading, renewable energy fluctuations
~\cite{Spatio-temporal_complexity}, \textit{etc.}
For example, energy trading happens at most of the time, inducing several
considerable local frequency deviations per day, even as four times per
hour.
Physically, the perturbations in a dynamical system can be regarded as
the initial states apart from the original stable equilibrium. Theoretical studies usually consider the distribution of power system
frequency fluctuations as uniform or normal, which are
not in line with the real situation.
The distribution of the frequency in realistic power systems~\cite{Non-Gaussian}
demonstrates the non-Gaussian characteristics of heavy tail and skewness,
which can be more accurately described by the Lévy-stable distribution.
Additionally, maximum fluctuations of frequency should be also set to an
appropriate value, otherwise the perturbations will be too small to
disturb the system or too large to be found in the realistic power systems.
Normally, the perturbations of frequency $\Delta f$ is bounded to
$\pm 1 \%$ to $\pm 4\%$ of rated frequency (50Hz or 60Hz)~\cite{Analysis_of_Power_System_Disturbances_Based_on}.
We set the perturbation limit of angular frequency as
$\Delta \omega = \Delta f = 20rad/s$. We define $m<N$ be the number of nodes simultaneously perturbed, where $m=1$ and $m>1$ refers to the single-node perturbations case and the multiple-node perturbations case, respectively.

\section{Temporal and Topological Embedding Deep Neural Network}
\label{section:GCN}

\begin{figure*}[!t]
	\centering
	\includegraphics[width = 18cm]{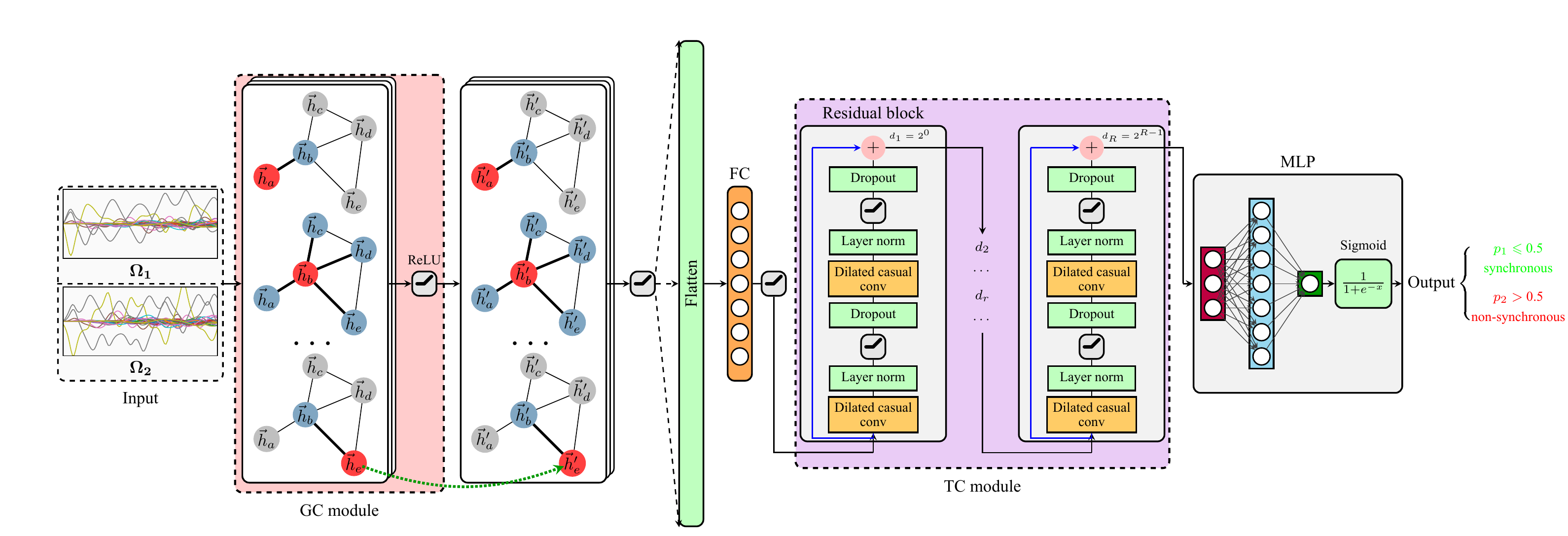}
	\caption{\textbf{Diagram of the TTEDNN structure.}}
	\label{fig:structure_RGCN}
\end{figure*}

Given perturbations and power network structures, the temporal and
topological features embedding in the transient dynamics in power systems
is crucial to the transient stability.
The TTEDNN model is proposed to predict the transient stability in power systems by extracting the spatial and temporal features
from the time series data generated with the transient dynamics effectively.
The TTEDNN structure is shown in Fig.~\ref{fig:structure_RGCN}.

\subsection{Data Representation}

The transient dynamics of power system is represented
as the multivariate time series data of angular frequency, donating as
$\boldsymbol{\Omega}_s =[ \boldsymbol{\omega}_1 ; \boldsymbol{\omega}_2 \; \cdots\; \boldsymbol{\omega}_i \; \cdots \; \boldsymbol{\omega}_N]^T$, 
where $N$ is the number of nodes in power systems,
$\boldsymbol{\omega}_i=[ \omega_0^i ; \omega_1^i ; \cdots ; \omega_j^i ; \cdots ; \omega_T^i ]$
is the time series of the frequency of node $i$ and $T$ is the length of time series. The label $y$ for $\boldsymbol{\Omega}_s$ is binary,
\textit{i.e.}, $y=1$ and $y=0$ for the initial states of $\boldsymbol{\Omega}_s$ will lead to
a stable steady-state and unstable steady-state, respectively.
In the TTEDNN prediction output gives the probability $p$ that the power
system will evolve to the stable steady-states or unstable steady-states.
Numerically, we take $p > 0.5$ for the stable steady-state and $p\leqslant 0.5$
for the unstable steady-state.

\subsection{Structure of TTEDNN}

The TTEDNN structure has three main parts, the
Graph Convolution (GC) modules, Temporal Convolution (TC) module and
Multi-Layer Perception (MLP) prediction layer. 

\textbf{GC Modules}.
TTEDNN starts with the $n$ GC modules to extract the topological features.
Each GC module is composed of GCN layer, Batch Normalization
(BN) layer and rectified linear unit (ReLU)~\cite{relu} activation function
sequentially.

The structure of the GCN layer can be represented as an undirected graph
~\cite{kipf2017semisupervised} $G=(\mathcal{V},\mathcal{E}, \mathbf{B})$,
where $\mathcal{V}\in \mathbb{R}^N$ is the set of neurons, 
$\mathcal{E}\in \mathbb{R}^E$ is the set of links between neurons and
$\mathbf{B} \in \mathbb{R}^{N \times N}$ is the adjacency matrix of
the graph.
The re-normalized adjacency matrix $\hat{\mathbf{B}}^{'}$ is often 
used in the GCN layer:

\begin{equation}
	\hat{\mathbf{B}}^{'}=\hat{\mathbf{D}}^{-1/2}\hat{\mathbf{B}}\hat{\mathbf{D}}^{-1/2}
	\label{eq:renormalize}
\end{equation}
where $\hat{\mathbf{B}}=\mathbf{B}+\mathbf{I_N}$ denotes the adjacency
matrix with self-loop, where $\mathbf{I_N}$ is the identity matrix.
$\hat{\mathbf{D}}$ represents the diagonal node degree matrix where 
$\hat{D}_{i,i}=\Sigma_{j}\hat{B}_{i,j}$.

The operation of the $i^{th}$ GC module is defined as

\begin{equation}
	\mathbf{H}^{i+1}=\sigma(BN(\hat{\mathbf{B}}^{'}\mathbf{H}^i \mathbf{W}^{i}+\mathbf{b}^i))
\end{equation}
where
$\sigma(\cdot)$ is the activation ReLU function,
$\mathbf{H}^{i+1} \in \mathbb{R}^{N \times C}$ denotes
the output states of $i^{th}$ GC module as well as the input
states of $(i+1)^{th}$ GC module, $\mathbf{W}^{i} \in \mathbb{R}^{C \times F}$
is the network weight of GCN layer, $\mathbf{b}^i \in \mathbb{R}^{N \times F}$
denotes bias.
Then the last $n^{th}$ GC module is connected to a flatten layer to
reshape the output states. 
Following the flatten layer, a fully-connected (FC) layer
is adopted to extract the topological features to feed into the TC module.

\textbf{TC Module}.
As the GC modules extracting the topological features, the TC module is
used to extract the temporal features.
As shown in Fig.~\ref{fig:structure_RGCN}, the TC module is composed of
$R$ residual blocks using 1D fully convolutional network (FCN)~\cite{DBLP:journals/corr/LongSD14}.
1D-FCN utilizes casual convolution technique and dilated convolution technique
as well as a residual connection.
The solely 1D-FCN structure produces an output with the same length of its
input, and the casual convolution technique ensures that the output emitted
by a 1D-FCN layer at time step $t$ is convolved only with elements from
time step $t$ and earlier in the previous layer.
Therefore, the TC module takes the whole history information into consideration
for the future prediction.
The disadvantage is that casual convolution technique can only look
back to the historical information with the size linear to the network
depth.
The casual convolution can be optimized with the dilated convolution technique
by introducing the exponential receptive field.
Therefore, the TC module can take all historical information into account
with smaller network depth.
Specifically, the dilated convolution operation $F(\cdot)$ can be defined
as a dilated transformation of a 1-D time series data $\mathbf{x}$:

\begin{equation}
	F(j) = \sum_{i = 0}^{k-1} f(i) \cdot x_{j-d \cdot i}
\end{equation}
where $f(i)$ is the convolution filter $f:{0,\ldots,k-1} \rightarrow \mathbb{R}$,
$k$ denotes the filter size, $d$ denotes the dilated factor and
$j=1,\cdots, n$ and $n$ is the size of the $\mathbf{x}$.
Adjusting dilation size can allow the top level of 1D-FCN represent a wider
range of the input as much as possible, thus expand the receptive field
extensively.
Moreover, residual connection is used to stabilize the network training
to make the layers to learn deep residual information as the modifications
to the identity mapping when the casual convolution is dilated with
considerable depth.
As a result, the $r$th residual block is defined as follows:

\begin{equation}
	\mathbf{z}^r = \sigma(\mathbf{x}^r+LF(F(\mathbf{x}^r)))
\end{equation}
where $\mathbf{x}^r$ and $\mathbf{z}^r$ denote the input and output of
the $r^{th}$ residual block respectively, $r=1,...,R$, $LF(\cdot)$ denotes
layer normalization technique~\cite{LN}.

\textbf{MLP Prediction Layer}.
Finally, MLP with the sigmoid activation function is utilized to generate
the prediction as a probability function.

\textbf{Grid-informed Adjacency Matrix}.
Taking the electrical and structural properties of power system into
consideration, three grid-informed adjacency matrices $\mathbf{B}$ are proposed
in the GC modules for the spatial features extraction.
First, according to~\cite{kipf2017semisupervised}, we choose the binary
adjacency matrix of power system added with self-connection, denoted as
$\mathbf{B}^{\uppercase\expandafter{\romannumeral1}}$ in Eq.~\ref{eq:coupling form 1}.
It considers the topology of power system, but ignores the weight of edges.
Second, active power flows are added as the weights of edges in the
adjacency matrix, denoted as $\mathbf{B}^{\uppercase\expandafter{\romannumeral2}}$
in Eq.~\ref{eq:coupling form 2}.
Third, we treat the maximum transmission capability of a transmission line
as off-diagonal elements in the adjacency matrix, together with active
power injections for diagonal ones, denoted as
$\mathbf{B}^{\uppercase\expandafter{\romannumeral3}}$ in Eq.~\ref{eq:coupling form 3}.

\begin{align}
	\label{eq:coupling form 1}
	B_{ij}^{\uppercase\expandafter{\romannumeral1}} & = \begin{cases}
		1 & (i,j) \in \mathcal{E} \ or \ i=j \\
		0 & (i,j) \notin \mathcal{E}         \\
	\end{cases} \\
	\label{eq:coupling form 2}
	B_{ij}^{\uppercase\expandafter{\romannumeral2}} & = \begin{cases}
		K_{ij} \sin(\delta_i - \delta_j) & (i,j) \in \mathcal{E}    \\
		0                                & (i,j) \notin \mathcal{E} \\
	\end{cases} \\
	\label{eq:coupling form 3}
	B_{ij}^{\uppercase\expandafter{\romannumeral3}} & = \begin{cases}
		K_{ij} & (i,j) \in \mathcal{E}                 \\
		0      & (i,j) \notin \mathcal{E},\, i \neq  j \\
		P_i    & i=j                                   \\
	\end{cases}
\end{align}

\textbf{Class Weighted Loss Function}.
In the training of the TTEDNN model, the class weighted binary cross entropy (BCE) is used as the loss
function with the $L2$ regularization defined as

\begin{equation}
	\begin{aligned}
		Loss= & \sum_{i}(\alpha_1 y_i \log p_i + \alpha_0 (1-y_i) \log (1-p_i)) \\
		& + \beta \sum_{k = 1}^{M} \frac{1}{2} (\|w_k\|^2 + \|b_k\|^2)
	\end{aligned}
	\label{eq:cost function}
\end{equation}
where $y_i$ denotes the label and $p_i$ denotes the model output of the
$i$th sample, $\alpha_0$ and $\alpha_1$ represent weight factors
corresponding to the stable state (frequency is synchronous at steady-states) and the unstable state (frequency is non-synchronous at steady-states), respectively.
Additionally, $w_k$ and $b_k$ serve as learnable network parameters,
$\beta$ is the regularization weight.

Class-weighted BCE is adopted for the loss function 
and proved significantly helpful for the training dataset with the great
imbalance of stability and instability.
The imbalance of the dataset results from the fact that the power systems
are globally stable under the most initial state perturbations.
There are more (less) stable (unstable) states samples in the dataset.


\section{Case Studies}
\label{section:case}

\subsection{Simulation Data}
\label{section:case.rawdata}

The transient dynamics simulation environment is set up
with the second-order power system model discussed in
Section~\ref{section:dynamic}. The simulation is solved by 4th order Runge-Kutta method with time step size $\Delta t=0.0125s$.
The IEEE 39-bus~\cite{Energy_function_book}
and IEEE 118-bus~\cite{Al-Roomi2015}
power systems are used for training and testing the TTEDNN model. The corresponding electrical parameters are obtained from MATPOWER 6.0 toolbox~\cite{matpower}. 

The training dataset contain the set of $\boldsymbol{\Omega}_s$ under the single-node perturbations.
The test dataset consists of the set of $\boldsymbol{\Omega}_s$ and the corresponding label $y$ under the single-node and multiple-node perturbations.
For the dataset under the
single-node perturbations, the initial states are randomly sampled
$\mathcal{K}_i$ times for all the nodes $i=1,\cdots N$ individually according to the uniform distribution of frequency.
As for the dataset under the $m$
multiple-node perturbations,  since there are too many choices of $m$
multiple nodes, $M$ combinations of $m$ multiple nodes are selected.
For each combination, the initial states of $m$ nodes are randomly sampled
$\mathcal{K}_m$ times according to the uniform distribution of frequency. For the dataset of single-node perturbations of the IEEE 39-bus power system, given
$\mathcal{K}_{i}=1000,\; i=1,\cdots 39$, 39000 samples in total are
generated with 35349 stable samples and 3651 unstable samples.
For the single-node perturbations dataset of the IEEE 118-bus system,
given $\mathcal{K}_{i}=441,\; i=1,\cdots 118$, 52038 samples in total are generated
with 50754 stable samples and 1284 unstable samples.
For the multiple-node perturbations dataset of the IEEE 39-bus power
system, given $m=3$, $M=60$ and $\mathcal{K}_m=1000$, 60000 samples in total are generated
with 43888 stable samples and 16112 unstable samples.
For the multiple-node perturbations dataset of the IEEE 118-bus power
system, given $m=3$, $M=60$ and $\mathcal{K}_m=200$, 12000 samples in total are
generated with 11132 stable samples and 868 unstable samples.
The dataset is divided into the training, validation and test datasets.  
60\% of the single-node dataset is used for the training, 20\% of the
single-node dataset is for the validation, and the remaining 20\% of the
single-node dataset and all the multiple-node dataset
are combined for the test dataset.

\subsection{Performance Analysis}

The four metrics, ACC, {Fall-out (FPR), Miss Rate (FNR), and
	AUC are used to measure the performance of the TTEDNN model, where
	$\mathrm{ACC = \frac{TP+TN}{TP+TN+FP+FN}}$, $\mathrm{FPR = \frac{FP}{FP+TN}}$,
	$\mathrm{FNR = \frac{FN}{FN+TP}}$, and $\mathrm{AUC = \int_{0}^{1} TPR \,dFPR}$
	where TP is the true positive, TN is the true negative, FP is the false
	positive, FN is false negative, ACC denotes the prediction accuracy
	rate, TPR denotes true positive rate, FPR denotes false positive rate,
	TNR denotes true negative rate and FNR denotes false negative rate.
	
	In order to understate the optimal length of time series, ACC and computational
	time are calculated with different length $T$ in Fig.~\ref{fig:trade_off}.
	ACC increases to the maximum around $T \approx 100$ while the computational
	time monotonically increases.
	Thus, $101$ is chosen to be the optimal length $T^*$ for the TTEDNN model. 
	
	\begin{figure}[!t]
		\centering
		\includegraphics[width = 8cm, keepaspectratio]{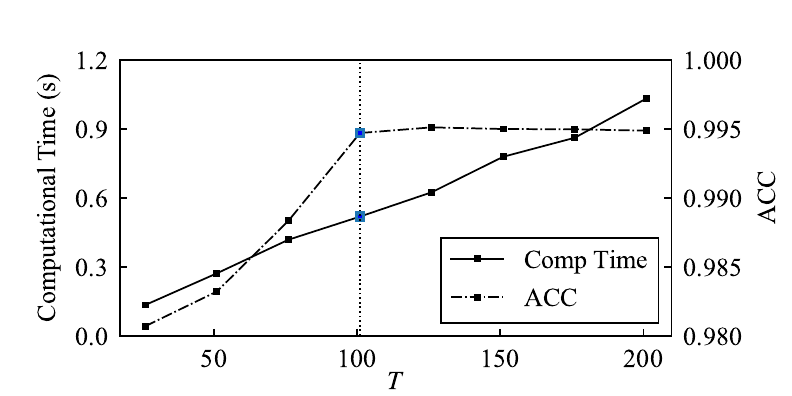}
		\caption{\textbf{ACC and
				computational time on the test dataset  for different time series length in the IEEE 39-bus power system.}
		}
		\label{fig:trade_off}
	\end{figure}
	
	The three grid-informed adjacency matrices in Eq.~\ref{eq:coupling form 1}
	to Eq.~\ref{eq:coupling form 3} are visualized in Fig.~\ref{fig:comparison_coupling_matrix}.
	The performance measurements in terms of the four metrics of ACC, TNR, TPR, and AUC are evaluated 
	for the three grid-informed adjacency matrices,
	$\mathbf{B}^{\uppercase\expandafter{\romannumeral1}}$, 
	$\mathbf{B}^{\uppercase\expandafter{\romannumeral2}}$, and
	$\mathbf{B}^{\uppercase\expandafter{\romannumeral3}}$, 
	respectively.
	The adjacency matrix $\mathbf{B}^{\uppercase\expandafter{\romannumeral2}}$
	shows the worst performance as shown in
	Fig.~\ref{fig:comparison_coupling_result}.
	The visualization in Fig.~\ref{fig:comparison_coupling_matrix} shows that
	the adjacency matrix $\mathbf{B}^{\uppercase\expandafter{\romannumeral2}}$
	is very sparse compared with the other two, which means that
	$\mathbf{B}^{\uppercase\expandafter{\romannumeral2}}$
	discards the useful information about the power system topology and
	electrical properties.
	The performance measurements of 
	$\mathbf{B}^{\uppercase\expandafter{\romannumeral1}}$
	and $\mathbf{B}^{\uppercase\expandafter{\romannumeral3}}$ are almost
	the same, while $\mathbf{B}^{\uppercase\expandafter{\romannumeral3}}$
	is slightly better on TPR.
	Hence, the grid-informed adjacency matrix
	$\mathbf{B}^{\uppercase\expandafter{\romannumeral3}}$ is used for the GC
	modules.
	
	\begin{figure}[!t]
		\centering
		\subfloat[Visualization of grid-informed adjacency matrices]{
			\includegraphics[width = 8cm, keepaspectratio]{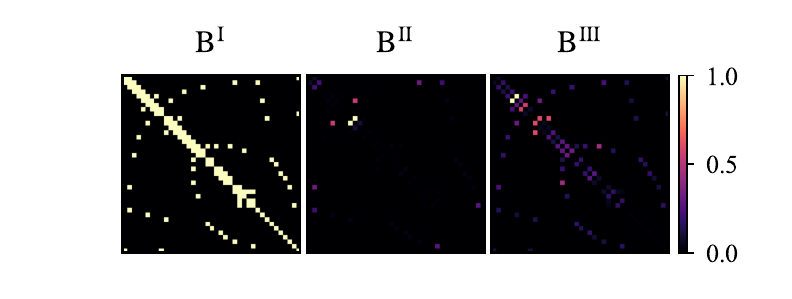}
			\label{fig:comparison_coupling_matrix}
		}
		\vspace{.0in}
		\subfloat[Performance in terms of ACC, TNR, TPR and AUC]{
			\includegraphics[width = 8cm, keepaspectratio]{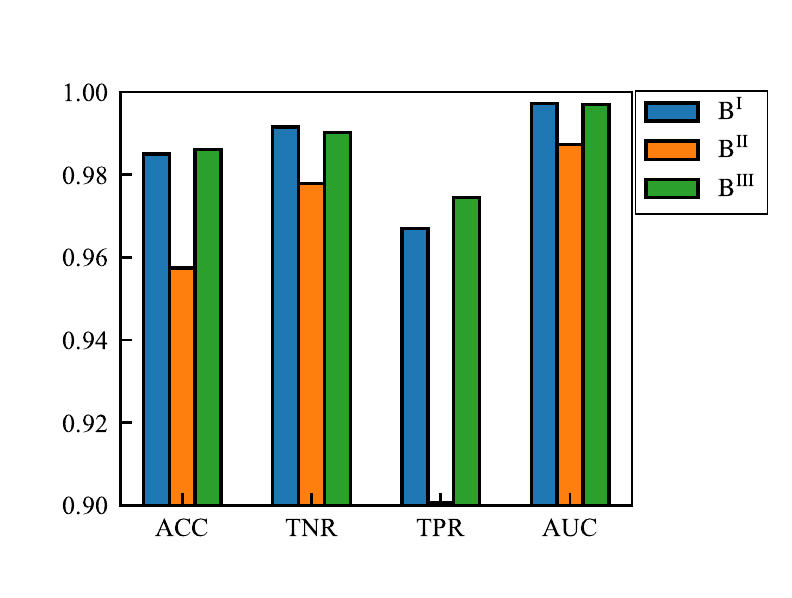}
			\label{fig:comparison_coupling_result}
		}
		\caption{\textbf{Visualization and performance comparison of three
				grid-informed adjacency matrices,}
			\protect\subref{fig:comparison_coupling_matrix} shows the
			visualization of three grid-informed adjacency
			matrices.
			\protect\subref{fig:comparison_coupling_result} shows performance
			comparison of three grid-informed adjacency
			matrices on the multiple-node test dataset.}
		\label{fig:comparison_coupling}
	\end{figure}
	We train and test the TTEDNN model using Tensorflow 2.3.1~\cite{tensorflow}
	and running on a server with Intel(R) Xeon(R) CPU E5-2620 v3.
	The TTEDNN model are tuned and validated according to the training score
	and validation score on the training and validation datasets.
	Two GC modules ($n=2$) with the same kernel size of $16$ are used to extract
	topological features.
	The dimension of the FC layer is (64, 1).
	The TC module has five residual blocks ($R=5$) and exponential dilated
	factors $d=2^{r-1}$ for $r=1,...,R$ with kernel size $k$=2 and the
	number of filters are $32$.
	The MLP prediction layer has the dimensions of (32, 1),
	(32, 1) for the input layer and the hidden layer respectively.
	The learning rate and batch size for the training are set to be
	$1e^{-3}$ and 256.
	$L_2$ regularization weight $\beta$ is set to $5e^{-4}$.
	Weight factor $\alpha_0$ is set to 1, and $\alpha_1$ is calculated on
	each batch as
	
	\begin{align}
		\alpha_1 =
		\begin{cases}
			256/\sum\limits_{i=1}^{256}y_i - 1 & \sum\limits_{i=1}^{256}y_i \neq  0 \\
			0                                  & \sum\limits_{i=1}^{256}y_i =  0.   \\
		\end{cases}
	\end{align}
	
	\subsection{Prediction Performance}
	
	\begin{table*}[!t]
		\centering
		\renewcommand\arraystretch{1.5}
		\caption{\textbf{Stability prediction under the single-node
				perturbations in the
				IEEE 39-Bus and the IEEE 118-Bus Power
				Systems.}}
		\label{table:single}
		\begin{threeparttable}
			\setlength{\tabcolsep}{2.8mm}{
				\begin{tabular}{ccccccccc}
					\toprule
					\multirow{2}{*}{Scenario} & \multicolumn{4}{c}{IEEE 39-bus} & \multicolumn{4}{c}{IEEE 118-bus}                                                                                \\
					\cmidrule(lr){2-5} \cmidrule(lr){6-9}
					& ACC(\%)                         & FNR(\%)                          & FPR(\%)       & AUC             & ACC(\%) & FNR(\%)       & FPR(\%) & AUC    \\
					\hline
					CNN                       & 98.36                           & 0.86                             & 9.33          & 0.9907          & 99.62   & 0.23          & 5.78    & 0.9928 \\
					\hline
					Attention-CNN             & 98.45                           & 0.97                             & 7.24          & 0.9930          & 99.69   & \textbf{0.14} & 6.64    & 0.9845 \\
					\hline
					GCN                       & 96.19                           & 3.33                             & 18.53         & 0.9558          & 99.25   & 0.18          & 2.15    & 0.9978 \\
					\hline
					GAT                       & 98.15                           & 2.20                             & 8.91          & 0.9852          & 99.53   & 0.18          & 6.14    & 0.9921 \\
					\hline
					NGCN                      & 84.27                           & 10.36                            & 12.17         & 0.8921          & 87.43   & 8.98          & 13.24   & 0.9135 \\
					\hline
					Proposed TTEDNN           & \textbf{99.38}                  & \textbf{0.65}                    & \textbf{0.36} & \textbf{0.9994}
					& \textbf{99.88}                  & 0.17                             & \textbf{0.00} & \textbf{0.9999}                                              \\
					\bottomrule
				\end{tabular}
			}
		\end{threeparttable}
	\end{table*}
	
	\begin{table*}[!t]
		\centering
		\renewcommand\arraystretch{1.5}
		\caption{\textbf{Stability prediction under the
				multiple-node
				perturbations in the
				IEEE 39-Bus and the IEEE 118-Bus Power
				Systems.}}
		\label{table:multiple}
		\begin{threeparttable}
			\setlength{\tabcolsep}{2.8mm}{
				\begin{tabular}{ccccccccc}
					\toprule 
					\multirow{2}{*}{Method} & \multicolumn{4}{c}{IEEE 39-bus} & \multicolumn{4}{c}{IEEE 118-bus}                                                                          \\
					\cmidrule(lr){2-5} \cmidrule(lr){6-9}
					& ACC(\%)                         & FNR(\%)                          & FPR(\%)       & AUC             & ACC(\%) & FNR(\%) & FPR(\%) & AUC    \\
					\hline
					CNN                     & 80.73                           & 11.93                            & 39.68         & 0.7976          & 98.18   & 0.34    & 20.74   & 0.9601 \\
					\hline
					Attention-CNN           & 82.49                           & 16.16                            & 21.27         & 0.7620          & 95.90   & 2.43    & 25.46   & 0.8942 \\
					\hline
					GCN                     & 93.21                           & 2.31                             & 10.45         & 0.9110          & 98.36   & 0.87    & 11.52   & 0.9814 \\
					\hline
					GAT                     & 90.26                           & 11.66                            & 4.41          & 0.9794          & 97.06   & 0.52    & 10.34   & 0.9956 \\
					\hline
					NGCN                    & 81.27                           & 15.56                            & 18.29         & 0.7821          & 86.62   & 13.21   & 16.43   & 0.8512 \\
					\hline
					Proposed TTEDNN         & \textbf{98.60}                  & \textbf{0.98}                    & \textbf{2.56} & \textbf{0.9970}
					& \textbf{99.08}                  & \textbf{0.09}                    & \textbf{9.33} & \textbf{0.9878}                                        \\
					\bottomrule
				\end{tabular}
			}
		\end{threeparttable}
	\end{table*}
	
	Fig.~\ref{fig:train_test_RGCN-TCN} shows the ACC and Loss change at the
	different training epochs with the validation dataset.
	It can be found that the training of the TTEDNN model converges quickly
	and smoothly.
	ACC increases sharply to 98\% within 20 epochs.
	\begin{figure}[!t]
		\centering
		\includegraphics[width = 8cm, keepaspectratio]{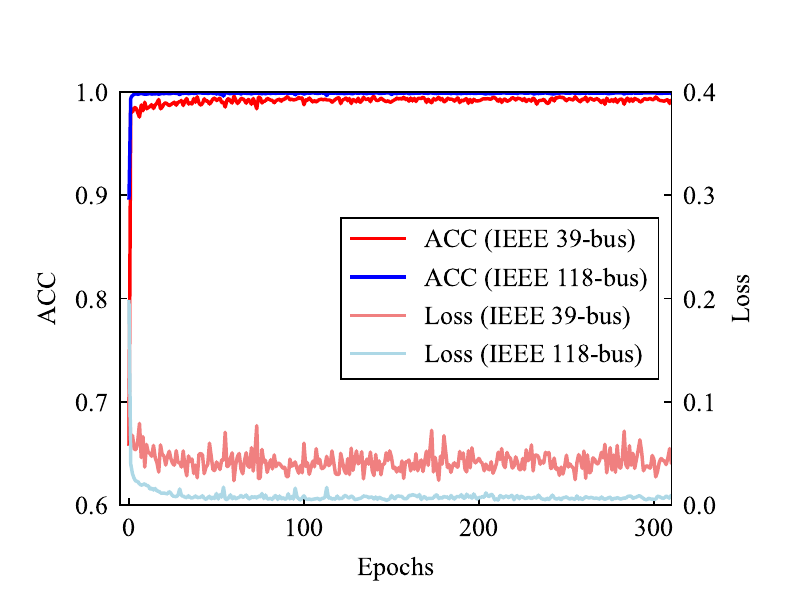}
		\caption{\textbf{ACC and class-weighted loss function $\it{v.s.}$ epochs.}
			The performance of the trained TTEDNN model on the validation dataset
			in terms of ACC and loss function for the IEEE 39-bus power
			system.}
		\label{fig:train_test_RGCN-TCN}
	\end{figure}
	
	Table~\ref{table:single} shows the performance of the transient prediction under the  single-node perturbations for the IEEE
	39-bus and IEEE 118-bus power systems.
	The test performance under the single-node test dataset shows that
	the TTEDNN model outperforms all the existing deep learning methods
	including CNN~\cite{CNN}, Attention-CNN~\cite{DBLP:journals/corr/abs-1807-06521},
	GCN~\cite{9369017}, Graph Attention Network (GAT)~\cite{GATTTT} and
	Node-Level GCN (NGCN)~\cite{Grattarola_2021}.
	The TTEDNN model has the best performance in terms of the ACC of 99.38\%
	and the AUC of 0.9994 for the IEEE 39-bus power system, and the ACC of
	99.88\% and the AUC of 0.9999 for IEEE 118-bus power system.
	The correct prediction of unstable states is the critically
	important in the practical implementation, thus we pay
	much attention to $FPR$, which represents the proportion of the correct prediction
	in all unstable samples.
	The TTEDNN model has the best FPR with only 0.36\%, \textit{i.e.},
	among all the 730 unstable samples, only three samples are mistaken predicted
	to be stable.
	Using the single-node test dataset of IEEE 118-bus power
	system, the attention-CNN method has the best FPR with 0.14\%, slightly
	better than the TTEDNN model with 0.17\%.
	
	The prediction performance on the multiple-node
	test dataset is also investigated since the
	multiple-node perturbations make the prediction task
	more complicated.
	Compared with the performance of the transient stability prediction
	under the single-node perturbations in Table~\ref{table:single}, 
	the TTEDNN model still has the best performance in Table~\ref{table:multiple}
	as the ACC of 98.60\% and the AUC of 0.9970 for the IEEE
	39-bus power system and the ACC of 99.08\% and the AUC of 0.9878
	for the IEEE 118-bus power system.
	However, the existing deep learning methods show 3\%-18\% drop in ACC and
	AUC, while the TTEDNN model only has very small changes in terms of ACC
	and AUC.
	Clearly the TTEDNN model is much more robust than the existing deep learning
	methods.
	
	The ACC in Table \ref{table:single} and Table \ref{table:multiple} indicates that the predictions of the TTEDNN model are highly close to the results of TDS method. Meanwhile, based on the Intel(R) Xeon(R) CPU
	E5-2620 v3 CPU, it takes approximately 0.5s and 5s for the trained TTEDNN model and the TDS method to forecast the transient stability, respectively, which indicates the TTEDNN model is 10 times faster than the TDS method.
	
	\section{Transfer Predictive Capability}
	\label{section:transfer}
	
	If the simulation can provide the sufficient information of the system
	behavior, the predictive model can learn from the simulation environment
	to make predictions.
	Learning from the simulation can be useful in the few-shot learning. 
	The availability of the transient dynamics data in the realistic
	power system is rare, particularly the data on the unstable transient
	dynamics.
	When the transient dynamics data in the realistic power system is missing,
	we can build the rigorous simulation
	environment of higher-order transient dynamics by the MATLAB PST toolbox
	~\cite{PST}.
	With the simulation environment, we can test the transfer capability of
	the TTEDNN model trained on the dataset generated with the coarse-grained
	simulation environment in Section~\ref{section:case.rawdata}. 
	The second-order power system model defined in Eq.~\ref{eq:swing_equation}
	can sufficiently describe the transient dynamics~\cite{PMID:24910217, power_book} of synchronous machines theoretically.
	The transfer capability of the TTEDNN model is worth investigation.
	The dataset generated with the rigorous simulation environment can be
	used to test the transfer capability of the TTEDNN model trained with
	the second-order simulation environment.
	
	We considered two type of loads, constant impedance and synchronous motors
	in the simulation environment, respectively.
	Table~\ref{tabel:TDS_comparsion} shows the transfer performance for the two
	types of loads.
	For the synchronous motor load, the trained TTEDNN model gives
	91.07\% ACC.
	For the constant impedance load, the performance of the TTEDNN model drops
	dramatically to 62.38\% ACC.
	Interestingly, the TTEDNN model has 100\% FPR while the FNR is less than
	15\%.
	In other words, the TTEDNN model makes the wrong prediction for the 85\%
	of the stable samples and correct prediction for all the unstable samples.
	Essentially, the TTEDNN model is a classifier extracting features of the input data for distinguishing the stable and
	unstable samples. The features from the hidden layer of the MLP in the TTEDNN model are visualized in Fig.~\ref{fig:hidden_layer} with the t-Distributed Stochastic Neighbor Embedding (t-SNE) dimensionality reduction
	technique~\cite{t-SNE}. The colored dots denote the value of classification
	confidence, \textit{i.e.},
	the sample is stable (unstable) when classification confidence is larger
	(smaller) than 0.5.
	
	Fig.~\ref{fig7a} and Fig.~\ref{fig7c} show the features distribution of
	training dataset labels and predictions, respectively.
	It can be found in Fig.~\ref{fig7a} that the TTEDNN model divide the stable
	and unstable samples into two clusters with a clear boundary.
	Clearly, the classification works well on the training dataset.
	However, as shown in Fig.~\ref{fig7b}, when the trained TTEDNN model is directly
	used for the samples generated with the simulation environment with the
	constant impedance loads, it fails to divide the stable and unstable samples.
	The classification results in a large part of stable samples that are
	mistakenly predicted to be unstable, as shown at the right corner of
	Fig.~\ref{fig7d}.
	Physically, the loads in the second-order power system model are
	equivalently synchronous motors in the power system model in the
	PST module.
	However, for the constant impedance, its physical characteristics are
	very different from the synchronous motors in nature, leading to distinct
	features distributions of the transient dynamics. 
	
	
	\begin{figure}[!t]
		\centering
		\subfloat[Training Data Labels]{
			\includegraphics[width = 4cm, keepaspectratio]{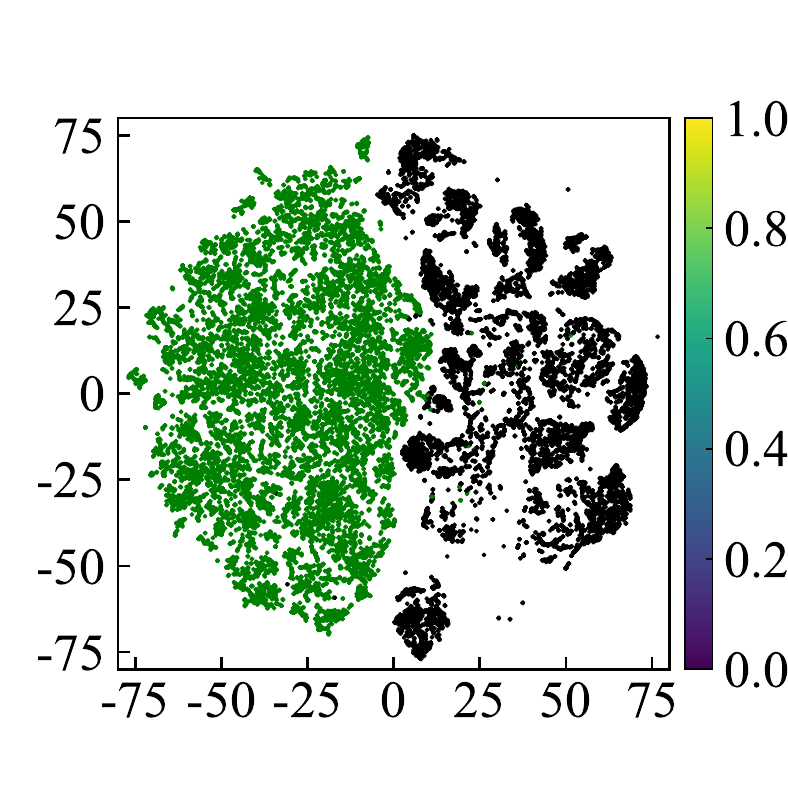}
			\label{fig7a}
		}
		\hspace{.0in}
		\subfloat[Transfer Test Data Labels]{
			\includegraphics[width = 4cm, keepaspectratio]{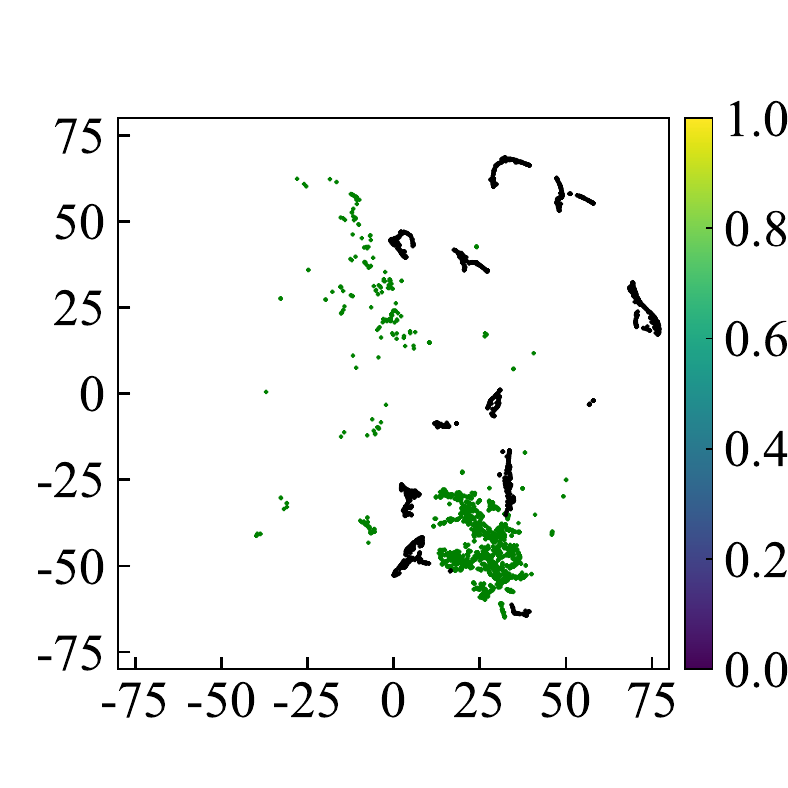}
			\label{fig7b}
		}
		\vspace{.0in}
		\subfloat[Division of Training Data]{
			\includegraphics[width = 4cm, keepaspectratio]{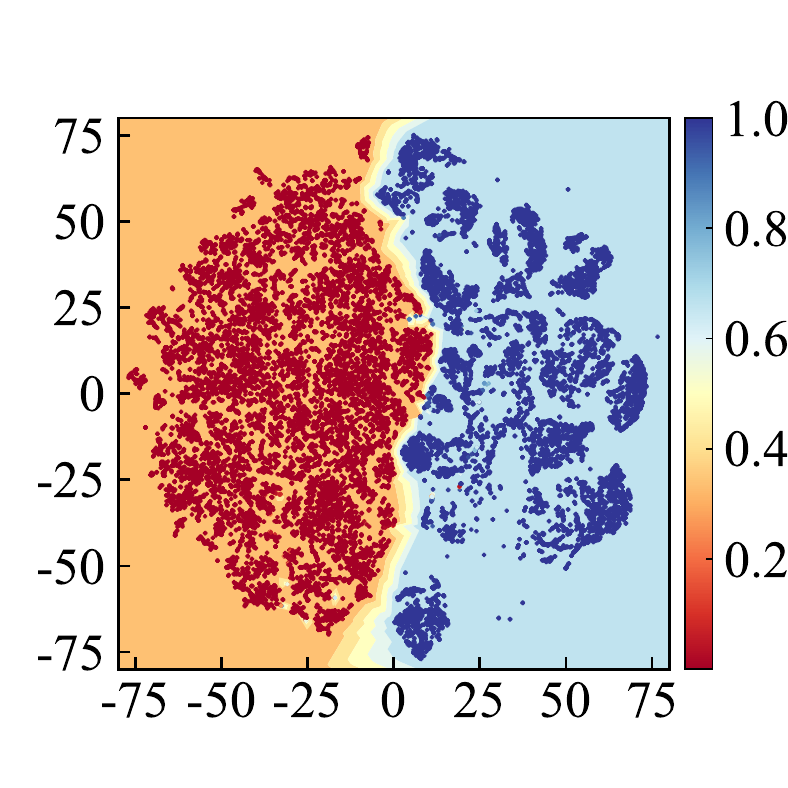}
			\label{fig7c}
		}
		\hspace{.0in}
		\subfloat[Division of Transfer Test Data]{
			\includegraphics[width = 4cm, keepaspectratio]{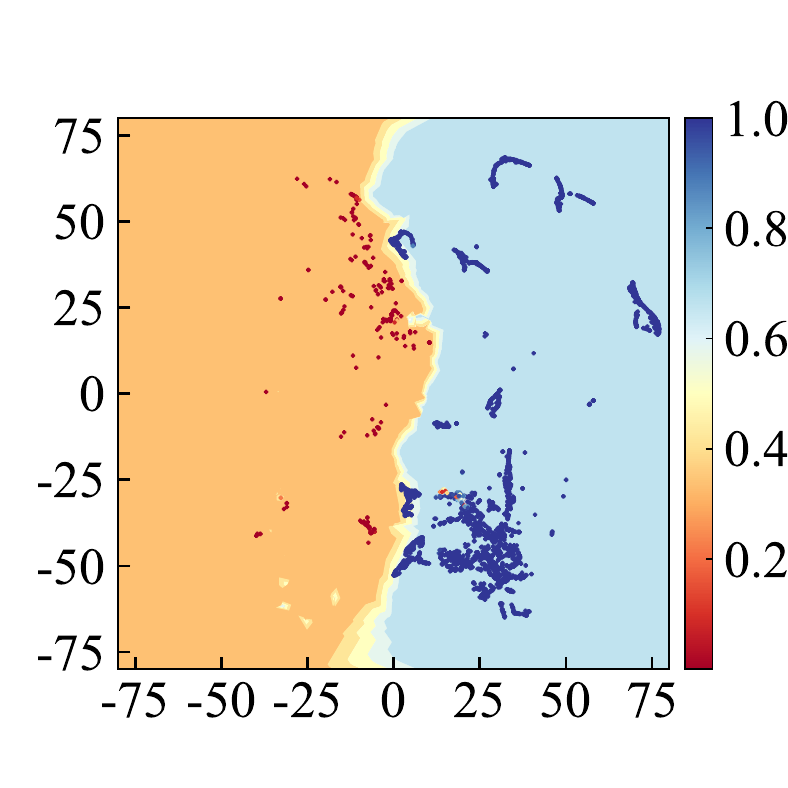}
			\label{fig7d}
		}
		\caption{\textbf{Visualization of high-dimensions features from the
				hidden layer with the constant impedance loads.}
			The \protect\subref{fig7a} and \protect\subref{fig7b} are training
			and transfer test data colored with labels,
			and \protect\subref{fig7c} and \protect\subref{fig7d} are training
			and test data colored with network outputs.}
		\label{fig:hidden_layer}
	\end{figure}
	
	\begin{table}[!t]
		\caption{\textbf{The transfer capability for  
				the IEEE 39-Bus Power System for the two types of the synchronous and constant impedance under
				single-node perturbations.}}
		\centering
		\setlength{\tabcolsep}{2.8mm}{
			\begin{tabular}{lcccc}
				\toprule 
				Scenario           & ACC(\%)         & FNR(\%)       & FPR(\%)       & AUC             \\
				\midrule 
				Constant impedance & 62.38           & \textbf{0.00} & 86.41         & 0.5001          \\
				Synchronous        & \textbf{91.07}  & 15.32         & \textbf{5.02} & \textbf{0.9707} \\
				\bottomrule 
			\end{tabular}
		}
		\label{tabel:TDS_comparsion}
	\end{table}
	
	\section{Concluding Remarks}
	\label{section:conclude}
	
	The TTEDNN model demonstrates robust and outstanding performance to
	predict the transient stability in power systems.
	First, the TTEDNN model maps the spatial information of power system
	topology into the GC modules as well as extracts the temporal and topological
	features from the transient dynamics of power systems with the GC and TC
	modules.
	Next, the TTEDNN model outperforms the existing deep learning models on
	almost all the performance metrics, \textit{e.g.}, it can reach
	99.38\% (98.60\%) ACC and 0.36\% (2.56\%) FPR under the single-node and multiple-node perturbations in the IEEE 39-bus power
	system.
	Third, the TTEDNN model is very efficient, much faster than the
	conventional TDS method.
	Fourth, the TTEDNN model has the transfer capability: the trained
	TTEDNN model with the second-order simulation environment can be directly used
	to predict the transient stability in the more realistic
	simulation environment based on the higher-order power system dynamics,
	especially when loads are synchronous motors.
	The transfer capability is extremely useful when the
	high-quality dataset is not available in the real-world power system.
	The TTEDNN model can be widely applied in transient stability analysis.
	The transfer learning capability of the TTEDNN model demands more theoretical
	and practical research in the future.

	\ifCLASSOPTIONcaptionsoff
	\newpage
	\fi
	
	\bibliographystyle{IEEEtran}   
	\bibliography{reference}
	
	
	
	
	
	
	
\end{document}